\documentclass[twocolumn,prl,showpacs,preprintnumbers,amsmath,amssymb]{revtex4}
\usepackage{graphicx}
\usepackage{dcolumn}
\usepackage{bm}
\usepackage{longtable}
\usepackage{epsfig}

\def\prl{Phys. Rev. Lett. }
\def\pra{Phys. Rev. A }
\def\prb{Phys. Rev. B }
\def\pre{Phys. Rev. E }
\def\jcis{J. Colloid Interface Sci. } 
\def\jap{J. Appl. Phys. }
\def\nimb{Nucl. Instrum. Methods Phys. Res. B }
\def\pnas{Proc. Natl. Acad. Sci. USA }
\def\jacs{J. Am. Chem. Soc. }

\begin{document}

\title{Observation of a Universal Aggregation Mechanism 
and a Possible Phase \\ Transition in Au Sputtered by Swift 
Heavy Ions}

\author{P. K. Kuiri,$^1$ B. Joseph,$^{1,}$\footnote{Present 
address: Istituto Tecnologie Avanzate, Contrada Milo 
91100, Trapani, Italy} H. P. Lenka,$^1$ G. Sahu,$^1$ J. 
Ghatak,$^1$D. Kanjilal,$^2$ and 
D. P. Mahapatra$^{1,}$\footnote{Electronic mail: 
dpm@iopb.res.in}}
\affiliation{$^1$Institute of Physics, Sachivalaya Marg, 
Bhubaneswar 751005, India\\ 
$^2$Inter University Accelerator Centre, Aruna Asaf Ali 
Marg, New Delhi 110067, India}

\begin{abstract}

Two exponents, $\delta$, for size distribution of $n$-atom 
clusters, $Y(n)\sim n^{-\delta}$, have been found in Au clusters 
sputtered from embedded Au nanoparticles under swift heavy ion 
irradiation. For small clusters, below 12.5 nm in size, $\delta$ 
has been found to be 3/2, which can be rationalized as occurring 
from a steady state aggregation process with size independent 
aggregation. For larger clusters, a $\delta$ value of 7/2 is 
suggested, which might come from a dynamical transition to 
another steady state where aggregation and evaporation rates 
are size dependent. In the present case, the observed decay 
exponents do not support any possibility of a thermodynamic 
liquid-gas type phase transition taking place, resulting in 
cluster formation.

\end{abstract}




\maketitle


The observation of emission of clusters of few atoms and molecules, 
from energetic ion impact was reported way back in 1958 
\cite{jap29}. This observation was quite surprising since the 
cluster binding energies are of the order of 1$-$2 eV, too 
small compared to the energies of the colliding ions. Since then 
there has been a lot of studies on the subject with an aim to 
understand the basic mechanisms behind cluster emission 
\cite{nimb164,prb66,prl87,nimb248,jap101,nimb244}. In most cases, 
involving low energy ion impact \cite{nimb164,prb66,prl87}, the 
cluster yield, $Y(n)$, as a function of number of constituent 
atoms, $n$, has been found to follow an inverse power law, 
$Y(n)\sim n^{-\delta}$, $\delta$ being a decay exponent. For 
small clusters with very few atoms, detected using time-of-flight, 
$\delta$ has been found to lie between 4 and 8 \cite{nimb164}.  
However, using transmission electron microscopy (TEM), it has 
recently been possible to study the size distribution of large 
clusters, collected on catcher foils, placed suitably during 
ion irradiations \cite{prl87,nimb248}. 

On the theoretical side there is a shock-wave model \cite{nimb21} 
where overlapping collision cascades, from low energy heavy ion 
impact, can result in the production of shock-waves in a medium. 
These can propagate to the surface. In some cases they can result 
in emission of a chunk of material with essentially the same 
atomic coordination as the target. This mechanism, yields a value 
of $\delta$ close to 2. Sputtering data from Au thin films, 
irradiated using four different ions with energies in the range of 
400 to 500 keV, have been shown to be in line with this \cite{prl87}. 
Competing with the above model, there is also a thermodynamic model 
\cite{nimb31} where the cascade of atomic displacements, produced 
in the near surface region of a target, can thermalize and expand 
into vacuum. The temperature is supposed to go beyond the 
liquid-gas critical temperature, $T_C$. Upon expansion and cooling 
the material can undergo a liquid-gas phase transition leading to 
cluster formation. In such a case the size distribution has been 
shown to follow a power law decay with a $\delta$ value close to 
7/3. 

As compared to continuous media (films or bulk), irradiation of nm 
sized metal islands or metal nanoparticles (NPs), embedded in a 
matrix, with a possibility of melting and evaporation, form a 
different class of systems. In this letter, we show that the size 
distribution of clusters emitted from such a system, under swift 
heavy ion (SHI) irradiation, falls under a universal class of 
aggregation. These systems possess non-equilibrium steady state 
solutions of mass distributions in the form of inverse power laws. 
In all such cases there is a competition between aggregation and 
breakup or evaporation, a delicate balance between the two leading 
to a variety of steady state mass distributions \cite{white}. 
Using a simple two-dimensional lattice model with jump between 
nearest sites and aggregation, Takayasu {\it et al.} \cite{pra37} 
have shown that the asymptotic distribution of mass or size always 
follows a power law only with the injection of a unit mass 
(monomer) at each site. Without the injection of monomers the 
solution, in the infinite time limit, corresponds to an aggregate 
with all the particles sticking together. Later the analysis was 
extended to include both positive and negative values for the 
dynamical variable which was taken to be {\it charge} rather than 
mass \cite{prl63}. The model included both injection and 
evaporation (through pair creation injection of unit positive and 
negative charges). The kinetics of aggregation were studied using
a {\it mean field} theory. The system was found to reach a steady 
state with a charge distribution following a power law with 
$\delta$ = 3/2 \cite{prl63}. This happens to be a very general 
case corresponding to a broad class of phenomena. As shown by 
Bonabeau {\it et al.} \cite{pnas96,pre51}, fish schools, with 
breakup and injection, show a similar aggregation, the size 
distribution showing a decay with $\delta$ = 3/2. This is seen 
even in economics related to distribution of wealth \cite{jpcs}. Such a 
result has also been obtained by Majumdar {\it et al.} \cite{prl81}
for aggregation in a mass conserving mean field type site-site 
interaction model. It has also been shown that small changes in 
the breakup parameters do not affect the decay exponent 
\cite{pre51,jpcs}. There is however a cutoff size which depends 
upon the competition between aggregation and breakup, and the 
time scales associated with them. 

Here we show, Au atoms sputtered by SHI from Au NPs follow a 
steady state aggregation as mentioned above. The observed size 
distribution is found to be in the form of a truncated power law 
with a $\delta$ value of 3/2. Beyond a critical size of about 12.5 
nm there is a drop off, which appears to be again in the form of 
another power law with a much larger exponent. Under the present 
irradiation conditions, the temperature of the NPs is known to 
go well above the vaporization temperature. But the size 
distribution shows power law decays with $\delta$ values of 3/2 
and 7/2, quite different from 7/3 as suggested for a liquid-gas 
type phase transition model \cite{nimb31}. Since the system 
indicates a steady state scenario there is no need to correct the 
size distribution against any breakup effects as applicable for 
cluster emission at lower irradiation energies \cite{prb71}. The 
results also indicate the thermal spike production from electronic 
energy loss to be an essential requirement for the present 
observations. 
 

For the present study, samples were prepared implanting 32 keV 
Au$^-$ ions to a fluence of 4$\times$10$^{16}$ cm$^{-2}$ into 
silica glass substrates followed by annealing at 850 $^\circ$C 
in air for 1 h. Hereafter, these will be called ``targets''. The 
Au implantation was carried out using a low energy negative ion 
implantation facility at the Institute of Physics (IOP), 
Bhubaneswar. For SHI, we have taken Au$^{8+}$ ions at 100 MeV. 
Three of the targets were irradiated with SHI at normal incidence 
to fluences F2, F5, and F10 with values of 2$\times 10^{13}$, 
5$\times 10^{13}$, and 1$\times 10^{14}$ ions cm$^{-2}$, 
respectively. These irradiations were carried out using the 16 MV 
Pelletron Accelerator at the Inter University Accelerator Centre 
(IUAC), New Delhi. For comparison, a fourth target was irradiated 
with 10 MeV Au$^{4+}$ ions to a fluence as given by F10 using the 
3 MV Pelletron accelerator at IOP, Bhubaneswar. In each case, both 
involving the low energy implantation (for target preparation) and 
the high energy target irradiations, the ion beams were raster 
scanned over an area of 1$\times$1 cm$^2$ for uniform irradiation. 
During the irradiation of the targets, sputtered particles were 
collected using catcher foils (in the form of carbon coated Cu 
grids usually used for TEM studies) placed at a distance of $\sim$ 
1 cm in front of the target at an angle of $\sim$ 15$^\circ$ with 
the sample surface. All the Au implantation and irradiations were 
carried out at room temperature. The starting target and the 
catcher foils with collected particles were imaged using a JEOL 
2010 UHR TEM operating at 200 kV. The Au content in the targets 
were checked before and after irradiation using Rutherford 
backscattering spectrometry (RBS) employing 1.35 MeV $^4$He$^+$ 
ions.

\begin{figure}[t]
\begin{center}
\includegraphics[width=8.25cm]{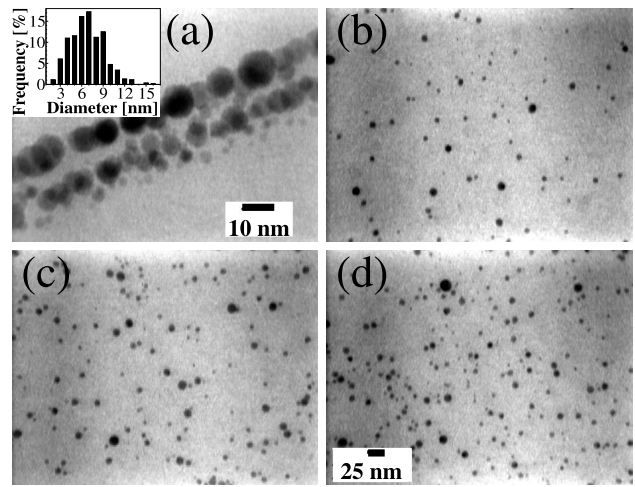}
\caption {(a) Bright-field cross-sectional TEM micrograph of the 
target showing Au NPs embedded in the matrix, before irradiation. 
Inset shows the size distribution of the Au NPs. (b)$-$(d) show 
plan-view TEM micrographs of the Au NPs on the catcher foils for 
SHI irradiations corresponding to the fluences of  
2$\times 10^{13}$, 5$\times 10^{13}$, and 1$\times 10^{14}$ 
ions cm$^{-2}$, respectively. Note that (b) and (c) are having 
same scale as (d).}
\label{tem}
\end{center}
\end{figure}

 
Bright-field cross-sectional TEM image [Fig \ref{tem} (a)] taken 
on the target shows a buried layer of spherical NPs with particle 
size decreasing from about 15 to 2 nm, progressively with depth. 
Planar TEM micrographs of the particles collected on the catcher 
foils following SHI irradiations to fluences of F2, F5, and F10 
are shown in Figs. \ref{tem}(b), (c), and (d), respectively. The 
larger particles are noticeably darker in the micrograph 
indicating them to be 3-dimensional entities. The particles are 
seen to have sizes ranging from about 1 to 20 nm. Figures 
\ref{tem}(b)$-$(d) also show an increase in NP number density 
with increase in fluence.

\begin{figure}[t]
\begin{center}
\includegraphics[width=8.cm]{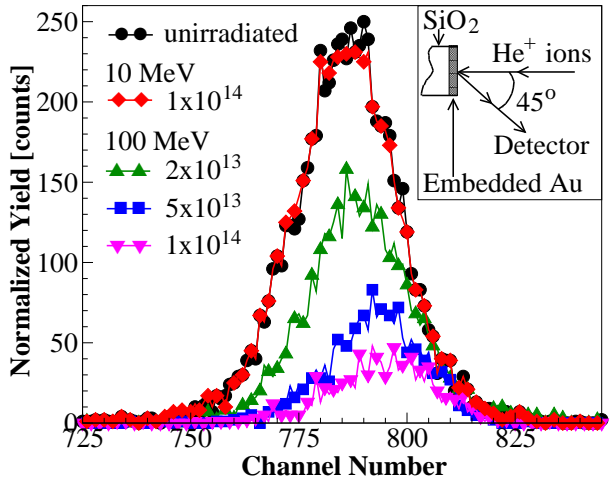}
\caption {(Color online) The Au part of the RBS spectra as 
measured on the targets before and after the irradiations. 
The irradiation fluences are in the unit of ions cm$^{-2}$. 
Inset shows the experimental arrangement.}
\label{rbs}
\end{center}
\end{figure}

From high resolution TEM and selected area electron diffraction 
measurements, the Au NPs in silica glass and on the catcher 
foils were found to be crystalline in nature with face centered 
cubic structure. Unlike the SHI irradiation no Au NP was found 
on the catcher foil for irradiation at 10 MeV. RBS measurements, 
carried out on the target after 10 MeV Au irradiation, showed 
no Au loss in contrast to what is found after the SHI 
irradiation  (Fig. \ref{rbs}). In the later case the Au loss 
is seen to increase with increase in fluence.

\begin{figure}[t]
\begin{center}
\includegraphics[width=8.cm]{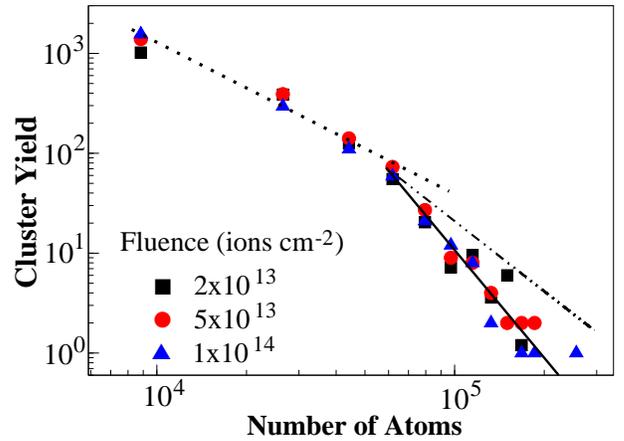}
\caption {(Color online) Size distribution in terms of the number 
of clusters of particular size (cluster yield) plotted against
the number of atoms in the cluster. Dotted, dot-dashed, and 
continuous lines correspond to power law decays with $\delta$ 
values of 3/2, 7/3, and 7/2, respectively. }
\label{sizedbn}
\end{center}  
\end{figure}

The size distributions of the NPs collected on the catcher 
foils, for various SHI fluences, are shown in Fig. \ref{sizedbn} 
where the number of observed Au NPs of different sizes, $Y(n)$, 
has been plotted against $n$, the number of atoms in the 
$n-$atom cluster or NP. To generate these data, many TEM 
micrographs of the same frame size [as shown in Figs. 
\ref{tem}(b)$-$\ref{tem}(d)] were analyzed taking each NP to be 
spherical in shape. The number of atoms in an $n-$atom NP is 
estimated multiplying the volume of the NP by the number density 
of atoms in bulk Au. A total of 1313, 2009, and 2047 particles 
were considered for SHI fluences of F2, F5, and F10 respectively.
Superimposed on the data is a straight line representing a power 
law distribution with a $\delta$ value of 3/2 (dotted line). 
This is seen to agree with the data very well up to a cutoff size 
of about 12.5 nm (corresponding to about 62000 atoms) beyond which 
evaporation or breakup effects are dominant. This region 
corresponding to larger clusters and is affected by fluctuations 
because of progressively small number of particles observed. For 
this region we have also shown a power law decay with a $\delta$ 
value of 7/2 (continuous line). The data points seem to follow 
this behavior. What is more important is that the data for three 
different fluences, taken on three different samples, show the 
same behavior. As shown earlier \cite{nimb256}, SHI irradiation 
results in a change in the size distribution of the embedded Au 
NPs. But this does not seem to affect the size distribution of 
the sputtered Au NPs. In the following section we present a 
discussion on various aspects of the observed phenomenon and the 
conditions under which such aggregation takes place. 


The first and formost requirement is the ejection of Au atoms from 
embedded Au NPs under SHI irradiation. This can happen due to the 
formation of localized inelastic thermal spikes \cite{nimb216} 
produced in the NPs resulting in their vaporization. In the 
inelastic thermal spikes model, passage of a SHI results in 
excitation and ionization of electrons in a cylindrical region 
around the ion path. This energy at first gets distributed into 
the electronic system through electron-electron interactions in 
a time scale of $\sim 10^{-13}$ s. Electron-lattice interactions 
cause a part of this energy to go to lattice atoms resulting in 
a temperature rise. In case of small NPs the temperature of the 
thermal spike may go well above that required for vaporization 
because of the small volume. There is also no dissipation of heat 
into the insulating surrounding matrix. Simulations indicate this 
to be true leading to vaporization of smaller particles 
\cite{prb67_22}, some of which get attached to other NPs leading 
to Ostwald ripening. Although not shown here, cross-sectional 
TEM images taken on the targets, after SHI irradiations at 
various fluences, do show this. Such a phenomenon does not seem 
to happen with 10 MeV Au ions where electronic stopping (2.5 keV 
nm$^{-1}$) is way below that at 100 MeV (13.5 keV nm$^{-1}$). 
The vaporized material must also come out of the matrix for the 
observed aggregation to take place. In silica glass SHI 
irradiation can result in a melting of a cylindrical zone around 
the beam path which, because of the pressure imbalance 
\cite{prb67_22}, can result in a squeezing out of the evaporated 
Au atoms \cite{nimb256}.

In such a case, it is not obvious that the expanding system could 
be in a thermalized state, with a temperature above a critical 
value, $T_C$, before going through a liquid-gas phase transition 
as proposed by Urbassek \cite{nimb31}. In case it goes through 
such a transition, through the liquid gas co-existence region, 
there would be droplet formation. The size distribution of the 
sputtered clusters is then expected to show a power law decay with 
a $\delta$ value as given by Fisher's critical exponent, $\tau$, 
which lies between 2 and 2.5. For a Van der Waal's gas the 
exponent $\tau$ has a value of 7/3. Compared to this, in the 
present case a $\delta$ value of 3/2 has been obtained for smaller 
clusters which changes over to 7/2 for larger ones. In fact there 
is no indication of an exponent close to 7/3 which is also shown 
in Fig. \ref{sizedbn} (dot-dashed line) for comparison. This rules 
out, any possible liquid-gas type phase transition taking place in 
the vaporized Au system resulting from SHI irradiation. On the 
other hand, exchange of particles between nucleation sites, within 
the framework of a mass-aggregation model, that takes diffusion, 
aggregation on contact, and dissociation, can result in a steady 
state aggregation process with a $\delta$ value of 3/2 \cite{prl81}.
 What is more interesting is that the exponent 3/2 appears in a 
variety of systems, constituting a universal class where mass plays 
the role of a control parameter. Mass conservation with steady 
injection of monomers is all that is required with other details 
of aggregation or breakup not being important. These results are 
also in disagreement with simulation results on cluster emission 
from Au NPs under self-ion irradiation at lower energies 
\cite{nimb180} where electronic stopping effects are neglected.


It is also important to understand the reason behind the cutoff 
observed in the cluster size distribution at about 12.5 nm and the 
change over taking place at that point. Such a cutoff can come 
from splitting of larger clusters \cite{pnas96}. But as long as 
aggregation and evaporation rates are independent of mass or size, 
the $\delta$ value, up to the cutoff, remains 3/2. On the other 
hand a change over can occur when the aggregation and evaporation 
rates become mass dependent. As shown by Vigil {\it et al.} 
\cite{prb88}, for a critical value of the ratio of the two rates, 
there can be a transition between a steady state mass 
distribution and geletion (infinite mass aggregate). A power law 
decay, with a $\delta$ value of 7/2, has been shown to occur at 
the transition point. It appears, in the present case, at cluster 
sizes greater than 12.5 nm, somehow both the aggregation and 
evaporation rates become mass dependent leading to a phase 
transition from one steady state behavior with a $\delta$ value 
of 3/2 to another with a $\delta$ value of 7/2. At the moment it 
is not clear as to how such a transition takes place.

However, based on the present results, it is difficult to rule 
out the occurrence of a liquid-gas type phase transition in 
sputtered material for any general ion-target 
combination where the ion energy also plays a crucial role. 
In fact some experimental data do exist in support of the 
above model \cite{jacs103}.
What has been shown here is the existence of a new mechanism of 
aggregation in sputtered particles, not shown earlier.


To conclude, swift heavy ion (100 MeV Au$^{8+}$) irradiation 
of Au NPs, embedded in silica glass, has been found to result 
in ejection of vaporized Au NPs which follow a rather universal 
aggregation mechanism occurring in nature. For smaller clusters,
the size distribution 
shows a power law decay with a $\delta$ value of 3/2 as observed 
in the steady state solution for non-equilibrium aggregation 
processes with a steady injection of monomers. For sizes greater
than 12.5 nm, there seems to be a change over to another steady 
state aggregation with a higher $\delta$ value indicating mass 
dependent effects coming into play. The results do not indicate 
any liquid-gas phase transition taking place in the sputtered
Au upon cooling.

The authors wish to thank S. M. Bhattacharjee and G. Tripathy 
of IOP, Bhubaneswar, for some very useful discussions and
suggestions. We are also thankful to the Pelletron group of the 
accelerator facilities at IUAC, New Delhi and IOP, Bhubaneswar, 
in particular Mrs. Ramarani Das, for the help provided during 
ion irradiation experiments. 
\vspace{-.5cm}

\end{document}